\documentclass[%
 reprint,
 amsmath,amssymb,
 aps,
]{revtex4-2}

\usepackage{graphicx}
\usepackage{dcolumn}
\usepackage{bm}
\usepackage{subfigure}

\begin{document}

\preprint{APS/123-QED}

\title{Complex unit lattice cell for low-emittance storage ring light source}

\author{Zhiliang Ren$^1$}%
\author{Zhenghe Bai$^{1,2}$}%
 \email{Corresponding author. \\baizhe@ustc.edu.cn}
\author{Penghui Yang$^1$}%
\author{Lin Wang$^1$}%
\author{Hongliang Xu$^1$}%

\affiliation{
    $^1$National Synchrotron Radiation Laboratory, University of Science and Technology of China, Hefei 230029, China
    }

\affiliation{
    $^2$Synchrotron SOLEIL, L'Orme des Merisiers, 91190 Saint-Aubin, France
    }

\date{\today}

\begin{abstract}
To achieve the true diffraction-limited emittance of a storage ring light source, such as $\sim$10 pm·rad for medium-energy electron beams, within a limited circumference, it is generally necessary to increase the number of bending magnets in a multi-bend achromat (MBA) lattice, as in the future upgrade plan of MAX IV with a 19BA replacing the current 7BA. However, this comes with extremely strong quadrupole and sextupole magnets and very limited space. The former can result in very small vacuum chambers, increasing the coupling impedance and thus enhancing the beam instabilities, and the latter can pose significant challenges in accommodating the necessary diagnostics and vacuum components. Inspired by the hybrid MBA lattice concept, in this paper we propose a new unit lattice concept called the complex unit lattice cell, which can reduce the magnet strengths and also save space. The complex unit cell is numerically studied using a simplified model. Then as an example, a 17BA lattice based on the complex unit cell concept is designed for a 3 GeV storage ring light source with a circumference of 537.6 m, which has a natural emittance of 19.3 pm·rad. This 17BA lattice is also compared with the 17BA lattice designed with conventional unit cells to showcase the benefits of the complex unit cell concept. This 17BA lattice also suggests a new type of MBA lattice, which we call the MBA lattice with semi-distributed chromatic correction.

\end{abstract}

\maketitle


\section{\label{sec:level1}Introduction}

The electron beam emittance of a storage ring can be written as~\cite{Borland2014Lattice}
\begin{equation}
\varepsilon\propto \frac{F}{J_{x}}\cdot\frac{E^{2}}{N_{b}^{3}},
\end{equation}
where $F$ is a lattice dependent factor, $J_x$ is the horizontal damping partition number, $E$ is the electron energy, and $N_b$  is the number of bending magnets (bends) in the ring. Based on this relation, there are two ways to reduce the emittance at a fixed energy $E$: (1) reducing $F/J_x$, and (2) increasing the number of bends $N_b$. The former depends on the property of the unit cells in the lattice. The unit cell with a combination of longitudinal gradient bend (LGB) and reverse bend (RB) can effectively decrease $F/J_x$, and it is possible to achieve an emittance even lower than the theoretical minimum emittance (TME) of a homogeneous-bend cell~\cite{Riemann2019low}. The latter leads to the multi-bend achromat (MBA) lattice concept~\cite{Einfeld2014First}, which is used in the design of fourth-generation storage ring light sources with emittances 1$\sim$2 orders of magnitude lower than those of third-generation light sources. 

Currently, fourth-generation light sources are being developed with the goal of achieving much lower emittances, approaching the diffraction limit for the X-ray wavelengths of interest. For a medium-energy fourth-generation light source, the true diffraction-limited emittance is $\sim$10 pm$\cdot$rad. In order to achieve such a low emittance within a limited ring circumference, it is necessary to increase the number of bends in the MBA lattice. For instance, a 19BA lattice for the future development of MAX IV was designed with a natural emittance of 16 pm·rad~\cite{Tavares2018future}, about 20 times lower than that of the present 7BA lattice. However, such an MBA lattice with many unit cells will result in extremely strong quadrupoles and sextupoles due to the very strong focusing and very low dispersion, and it will also result in very tight spaces between magnets. A study in Ref.~\cite{Borland2014Lattice} showed that, for a fixed ring circumference consisting of $N_b$ identical unit cells (one bend in one unit cell), as the number of cells increases, the strengths of quadrupoles and sextupoles also increase as
\begin{equation}
S_{quad.}\propto \sim{N_b}^{2},    S_{sext.}\propto \sim{N_b}^{3}.
\end{equation}
In this situation, from Eqs. (1) and (2), we can know that the sextupole strengths are approximately inversely proportional to the emittance. Stronger magnets will result in smaller vacuum chambers, which can increase the coupling impedance and thus enhance the beam instabilities.

The hybrid MBA lattice concept~\cite{Biasci2014low} provides us with inspiration for reducing magnet strengths and saving space in the design of low-emittance lattices. Unlike the conventional MBA lattices~\cite{Tavares2014max,Streun2018sls}, which use distributed chromatic correction and place sextupoles in each unit cell, the hybrid MBA lattice only locates sextupoles in one pair of dispersion bumps. As a result, the hybrid MBA lattice requires a reduced number of magnets and significantly weaker sextupoles in comparison to the conventional MBA lattice. For a given lattice cell length, if fewer magnets are needed, more available space can be used to accommodate necessary diagnostics and vacuum components. From a conventional perspective, the hybrid 7BA lattice is composed of two matching sections and five unit cells. While from a novel perspective, we can regard the five unit cells as a complex unit cell with sextupoles on both sides. Thus the hybrid 7BA lattice can also be seen as being composed of two matching sections and one complex unit cell, similar to a triple-bend achromat lattice. With this novel perspective, we propose a new unit lattice cell concept in this paper, called the complex unit lattice cell, for reducing the strengths of multipole magnets and also saving space. We note that Ref.~\cite{Wang2018Complex} proposed the complex bend concept, which consists of a number of small dipoles interleaved with strong focusing and defocusing quadrupoles. Compared to the complex bend concept, the complex unit cell concept has a wider scope, since a complex bend can be one part of a complex unit cell. 

This paper is organized as follows. In Section II, we first introduce the concept of the complex unit cell, and use a simplified model to study the properties of the cell, and then a practical example of the complex unit cell is given and compared with the conventional unit cell. Section III presents an MBA lattice with complex unit cells, which is also compared with the MBA lattice with conventional unit cells. Further study is given in Section IV, and finally, conclusion is given in Section V.

\section{Complex Unit Lattice Cell}

\subsection{\label{sec:level2}Concept}

\begin{figure}
  \centering
      \includegraphics[width=0.45\textwidth]{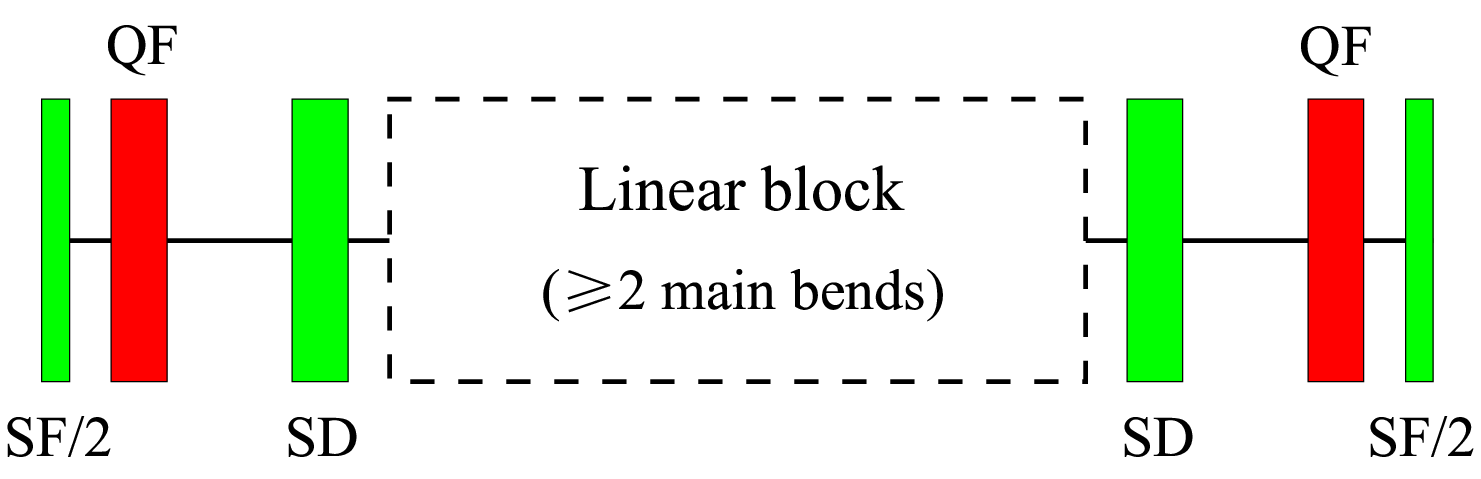}
  \caption{Schematic of a complex unit lattice cell. The dashed block represents a combination of linear magnets and drifts, with $\geqslant 2$ main bends. Outside the block, the red magnets, QF, are horizontally focusing quadrupoles; and the green magnets, SF and SD, are horizontally focusing and defocusing sextupoles, respectively.}
  \label{fig1}
\end{figure}

The schematic of a complex unit lattice cell is shown in Fig.~\ref{fig1}. The central part of the cell, shown as a dashed block, is a combination of linear magnets and drifts, with $\geqslant 2$ main bends and no nonlinear magnets. The linear magnets include quadrupoles and various bends, such as combined-function bends (CBs), LGBs and RBs. The sextupoles, which are placed only on both sides of the cell for correcting chromaticities, are shared by the central part. If the central part has only one main bend, then the cell becomes a conventional unit cell. Since there is no chromaticity correction in the central part of the cell, fewer nonlinear magnets are needed compared to conventional unit cells with the same number of main bends. This allows the complex unit cell to provide more space for accommodating other essential components.

The dispersion function after a drift, $\eta_1$, is given by $\eta_1$=$\eta_0+\eta_0^\prime\cdot l_D$, where $\eta_0$ and $\eta_0^\prime$ are the initial dispersion and its derivative, respectively, and $l_D$ is the length of the drift. Longer drift and higher $\eta_0^\prime$ help to obtain a larger dispersion. Due to the fewer magnets used, a complex unit cell can have longer drifts on both sides, enabling the creation of high dispersion regions, just like in the hybrid MBA lattice. Since the chromaticity compensation provided by sextupoles is proportional to the dispersion at their locations, a complex unit cell can require relatively weak sextupole strengths. Additionally, similar to the hybrid MBA lattice, the quadrupoles in the central part of the complex unit cell could have very strong strengths, but we can reduce their strengths by increasing their lengths due to the more space available in the cell.

Specifically, the strengths of sextupoles are determined not only by the dispersion at their locations, but also by the beta functions at their locations and the chromaticities that need to be corrected. Due to the complexity of the complex unit cell, we will numerically study the properties of the cell using a simplified model.

\subsection{\label{sec:level2}Study with a simplified model}

\begin{figure*}
    \centering
    \subfigure[Case 1]{
       \includegraphics[width=0.40\textwidth]{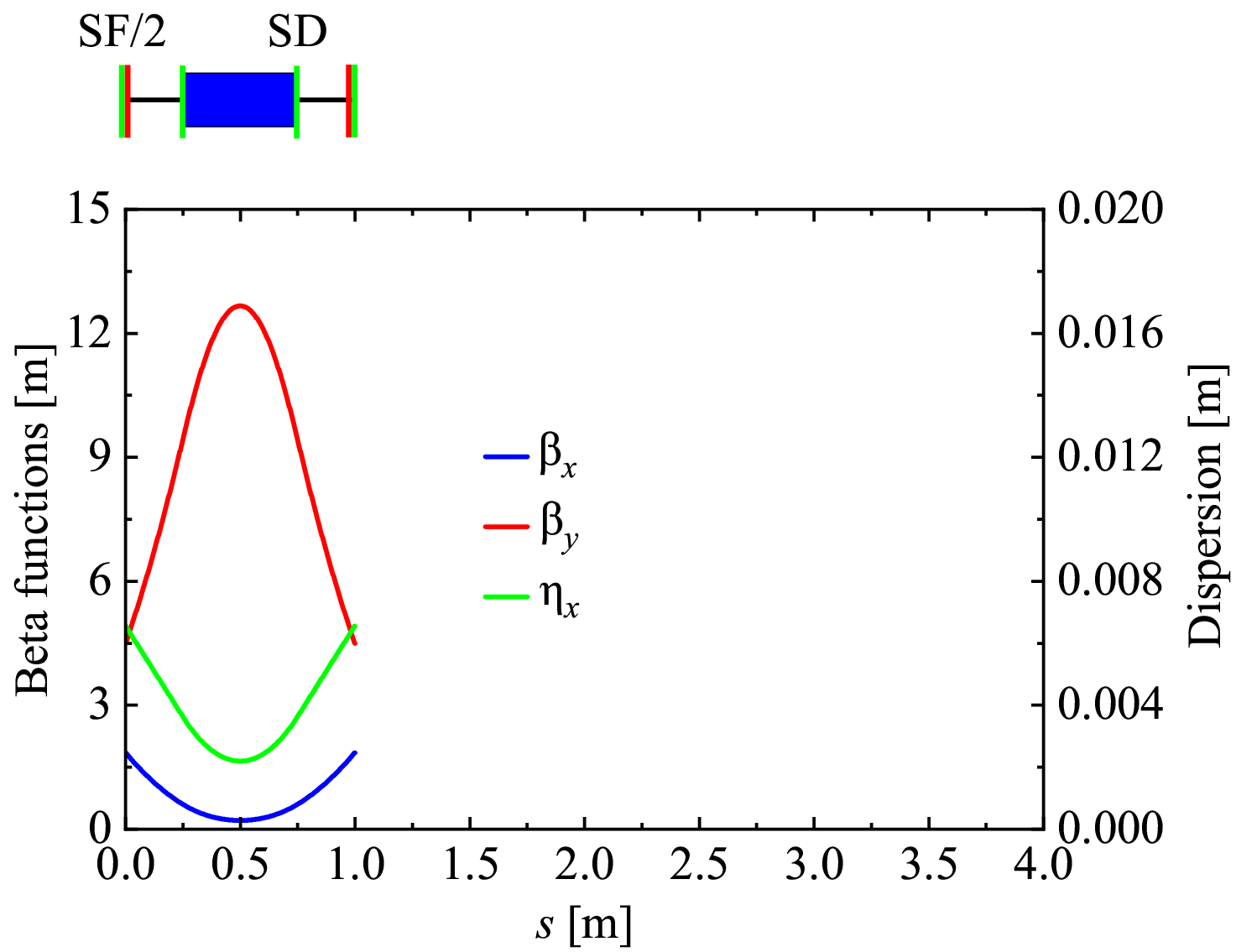}
    \label{fig201}
    }
    \hspace{0.05\linewidth}
    \subfigure[Case 2]{
       \includegraphics[width=0.40\textwidth]{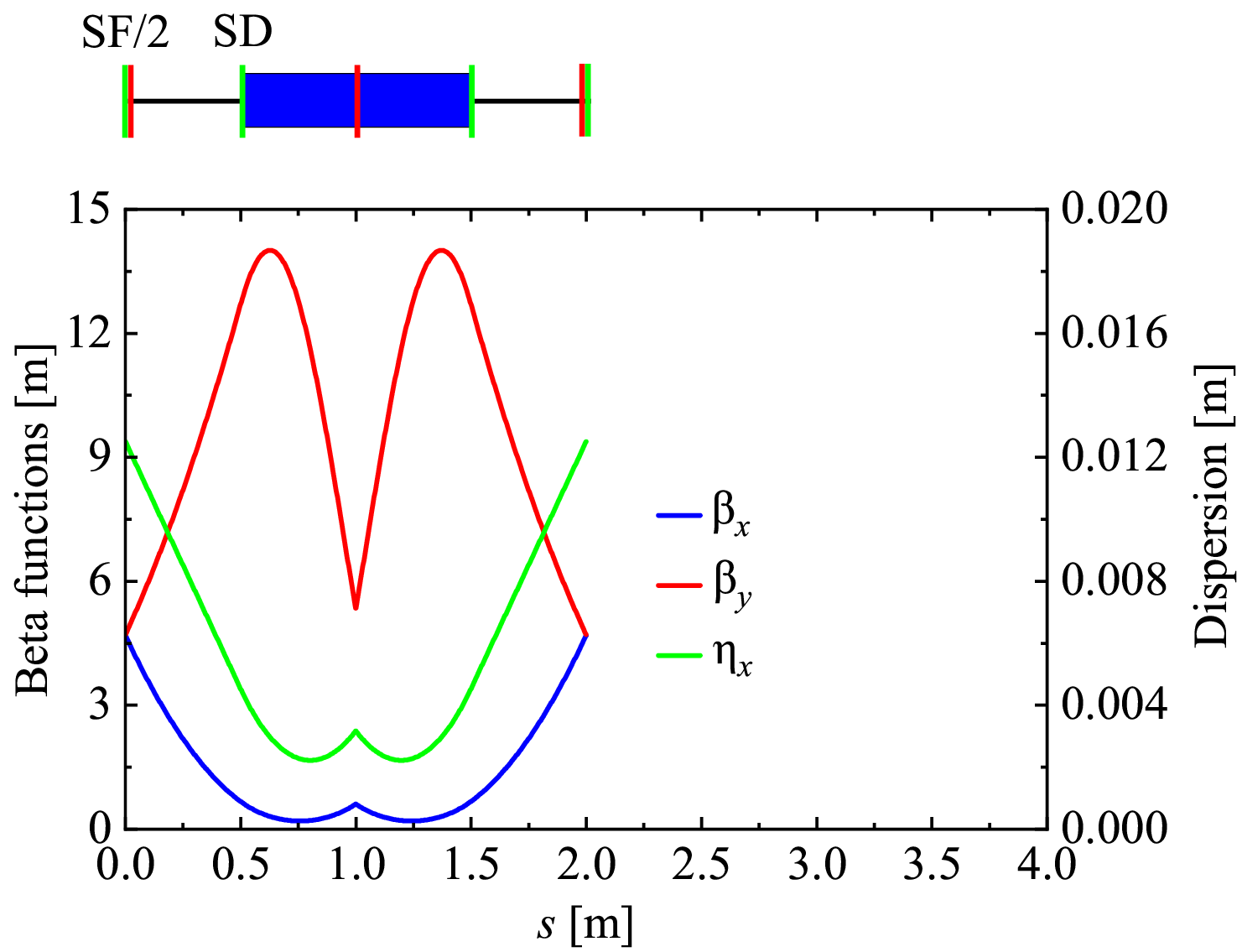}
    \label{fig202}
    }
    \subfigure[Case 3]{
       \includegraphics[width=0.40\textwidth]{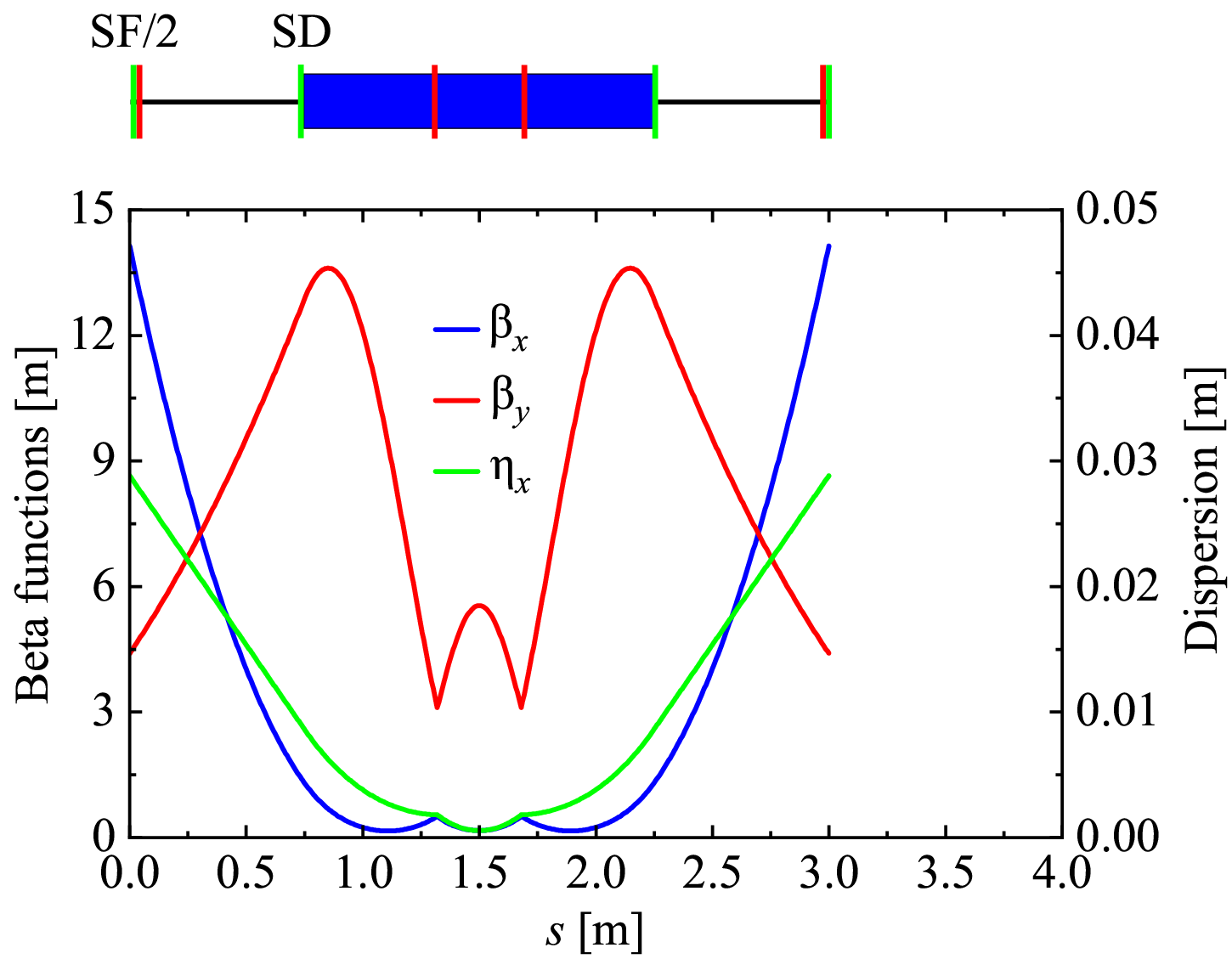}
    \label{fig203}
    }
    \hspace{0.05\linewidth}
    \subfigure[Case 4]{
       \includegraphics[width=0.40\textwidth]{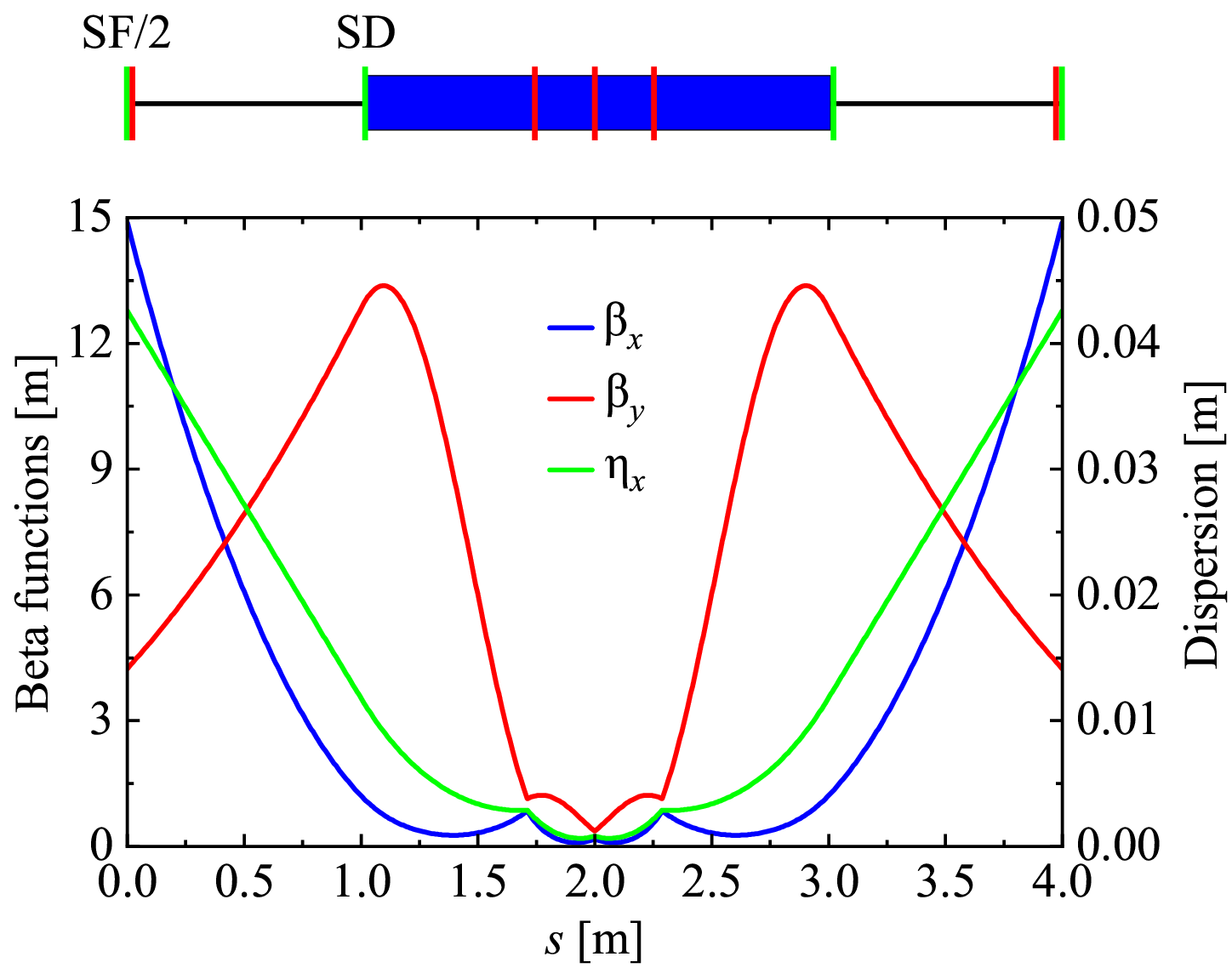}
    \label{fig204}
    }
\caption{\label{fig2} Four lattice solutions with symmetric optics for the conventional unit cell (a), and three complex unit cells (b), (c) and (d) with the number of bends $n=$ 2, 3, 4, respectively. In the magnet layout of each case, bends are shown in blue, and quadrupoles in red and sextupoles in green.}
\end{figure*}

To simplify the study of the complex unit cell, we adopted CBs as the main bends, and used the thin lens approximation for the quadrupoles. The central part of the cell, i.e. the dashed block in Fig.~\ref{fig1}, is simply composed of $n$ CBs ($n \geqslant 2$) interleaved with $n-1$ focusing quadrupoles, with no drift between them. On both ends of the cell, there are two focusing quadrupoles. Besides, we also considered sextupoles to correct the chromaticities in the cell, which are also treated as thin lenses. A focusing sextupole is located directly adjacent to the quadrupoles on both ends, and two defocusing sextupoles are directly adjacent to the central part. All of the magnets are symmetric with respect to the middle plane of the cell. One can see the magnet layout of such a cell model in Fig.~\ref{fig2}.

Here we study complex unit cells with varying numbers of bends ($n=$ 2, 3, 4), as well as the conventional unit cell with a single bend for comparison. Figure~\ref{fig2} shows the conventional unit cell and three complex unit cells with $n=$ 2, 3, 4, which are denoted as Case 1, Case 2, Case 3 and Case 4, respectively. In order to make a fair comparison between these unit cells, we set the following conditions: (1) the bending angle of each unit cell is proportional to its number of bends, (2) the length of each unit cell is also proportional to its number of bends, and (3) the total length of drifts in each unit cell is set to half of the unit cell length. Then some parameters need to be specified for the numerical calculation. The beam energy is set to 3.0 GeV. The bending angles of Case 1, Case 2, Case 3 and Case 4 are 1°, 2°, 3° and 4°, respectively; and their cell lengths are 1 m, 2 m, 3 m and 4 m, respectively. To obtain reasonable solutions, we set the following constraints: (1) the beta functions $\beta_{x,y} <$ 15 m, and (2) $J_x <$ 2. 

\begin{figure}
  \centering
      \includegraphics[width=0.45\textwidth]{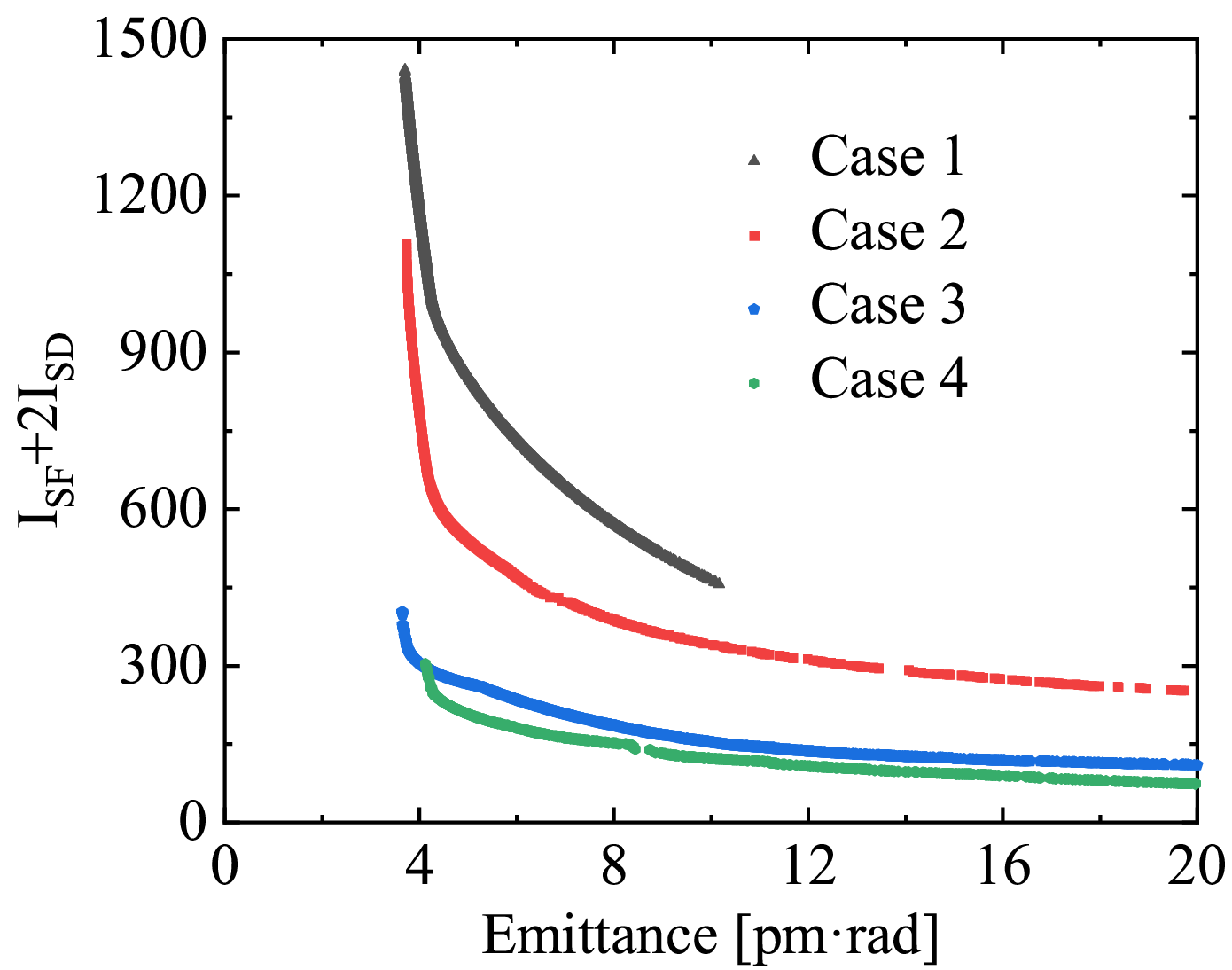}
  \caption{Pareto fronts of natural emittance and total integrated strength of sextupoles for the conventional unit cell and three complex unit cells.}
  \label{fig3}
\end{figure}

For each unit cell case, we minimized two objectives: (1) the natural emittance, and (2) the sum of absolute integrated strengths of sextupoles in one unit cell. For the calculation of sextupole strengths, the chromaticities were corrected to (0, 0) in each case. The optimization was performed using a multi-objective genetic algorithm (MOGA)~\cite{Yang2009global,Yang2011multiobjective} with the strengths of bends and quadrupoles and the lengths of bends as variables. MOGA was run with a population of 15000 and 100 generations for each case, and the Pareto fronts obtained are shown in Fig.~\ref{fig3}.

From Fig.~\ref{fig3}, we clearly see that complex unit cells can achieve the same emittance with much weaker sextupole strengths compared to the conventional unit cell. Due to the maximum beta functions being constrained, the minimum emittance of Case 4 is slightly larger than those of the other three cases. For the four cases, as $n$ increases, the sextupole strengths decrease due to the dispersion in the drifts becoming larger. However, the difference in sextupole strengths between Case 3 and Case 4 is not as large as expected, which is due to that the natural chromaticities of Case 4 are larger. In addition, the strength difference between Case 2 and Case 3 is larger. This is because starting from Case 3, the bends in a complex unit cell begin to form a longitudinal gradient on a macroscopic scale, similar to a LGB, which helps to increase the dispersion, thus reducing the sextupole strengths. The four unit cell solutions presented in Fig.~\ref{fig2} were taken from the four Pareto fronts of Fig.~\ref{fig3}, which have almost the same emittance of about 6 pm·rad. From Fig.~\ref{fig2}, it can be seen that the maximum dispersion value in Case 2 is about twice that of Case 1, while the maximum dispersion in Case 3 is 4$\sim$5 times that of Case 1.

Compared to the conventional unit cell, complex unit cells have more magnet families for optical focusing, which means more knobs can be used for tuning optics. It is easy to know that, for the conventional unit cell in Fig.~\ref{fig2}, if the horizontal and vertical cell tunes are fixed, the linear optics will also be fixed. While for complex unit cells, the optics are not unique for fixed cell tunes. Since resonance driving terms~\cite{bengtsson1997sextupole} are related to linear optics, the tunability of the optics of a complex unit cell can be used to some extent for nonlinear dynamics optimization.

\begin{figure} 
  \centering
      \includegraphics[width=0.45\textwidth]{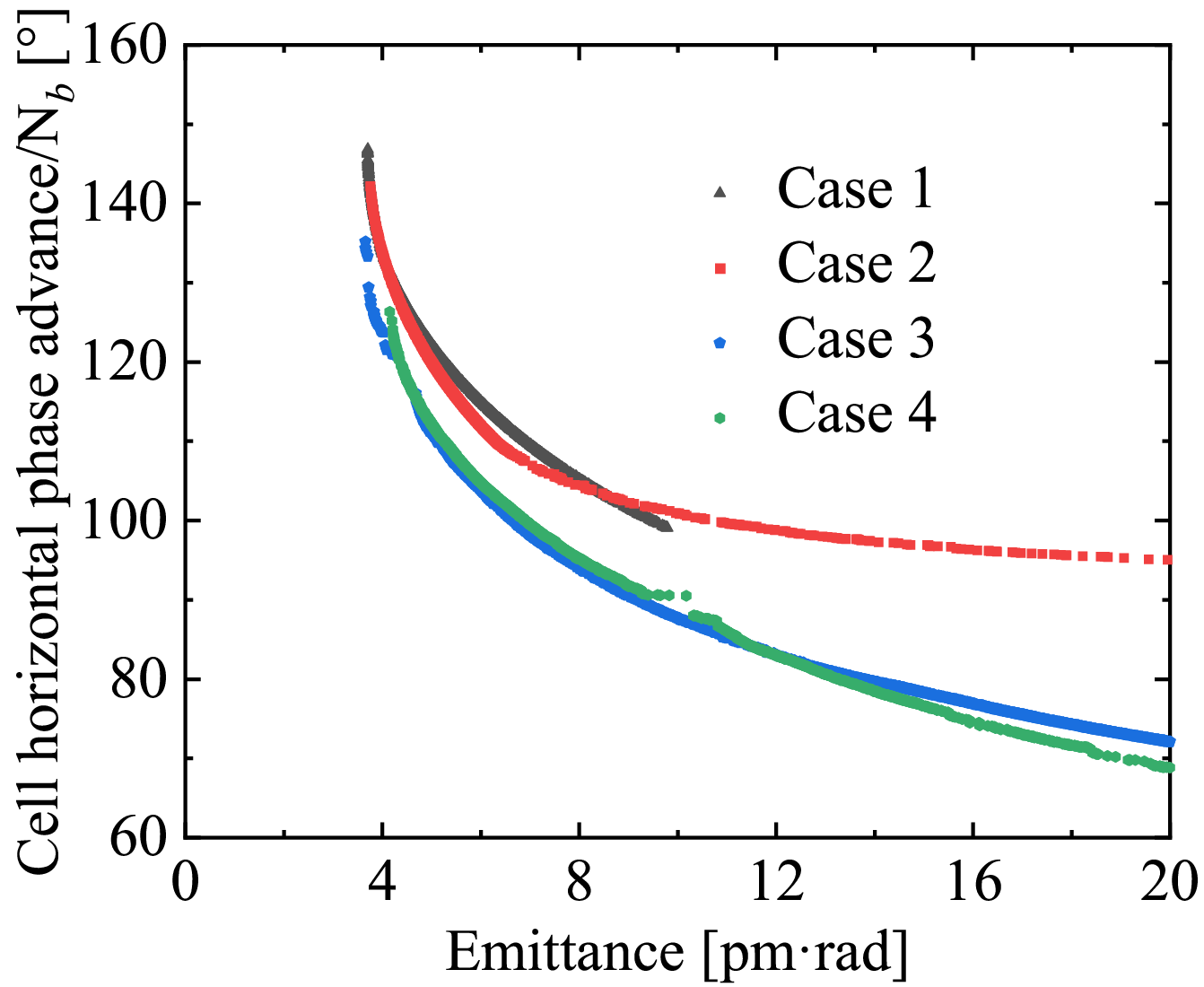}
  \caption{Pareto fronts of natural emittance and horizontal cell tune divided by the number of bends $n$ for the conventional unit cell and three complex unit cells.}
  \label{fig4}
\end{figure}

Also using the MOGA with the same settings as the previous optimization, we minimized: (1) the natural emittance and (2) the horizontal cell tune divided by $n$, denoted as $\nu_x/n$, for each unit cell case. The Pareto fronts obtained are shown in Fig.~\ref{fig4}. The Pareto front of the conventional unit cell, Case 1, truncates at a horizontal cell tune of 0.275 due to the constraint of $J_x <$ 2. As the horizontal cell tune becomes smaller, $J_x$ becomes larger. However, complex unit cells can have wider ranges of $\nu_x/n$, which reflects the better tunability of the optics of complex unit cells. The complex unit cells, Case 3 and Case 4, can obtain lower emittances at the same tune $\nu_x/n$. This is due to their higher ratios of the horizontal phase advance of the bend section to that of the whole cell. Note that Case 3 and Case 4 have larger horizontal beta functions in the straight sections, which means the horizontal phase advances of the straight sections are smaller. There are breakpoints on the Pareto fronts of Case 3 and Case 4, corresponding to unstable solutions with the horizontal cell tunes $\nu_x$ equal to 1.

\subsection{\label{sec:level2}A practical example}
\begin{figure} 
  \centering
      \includegraphics[width=0.45\textwidth]{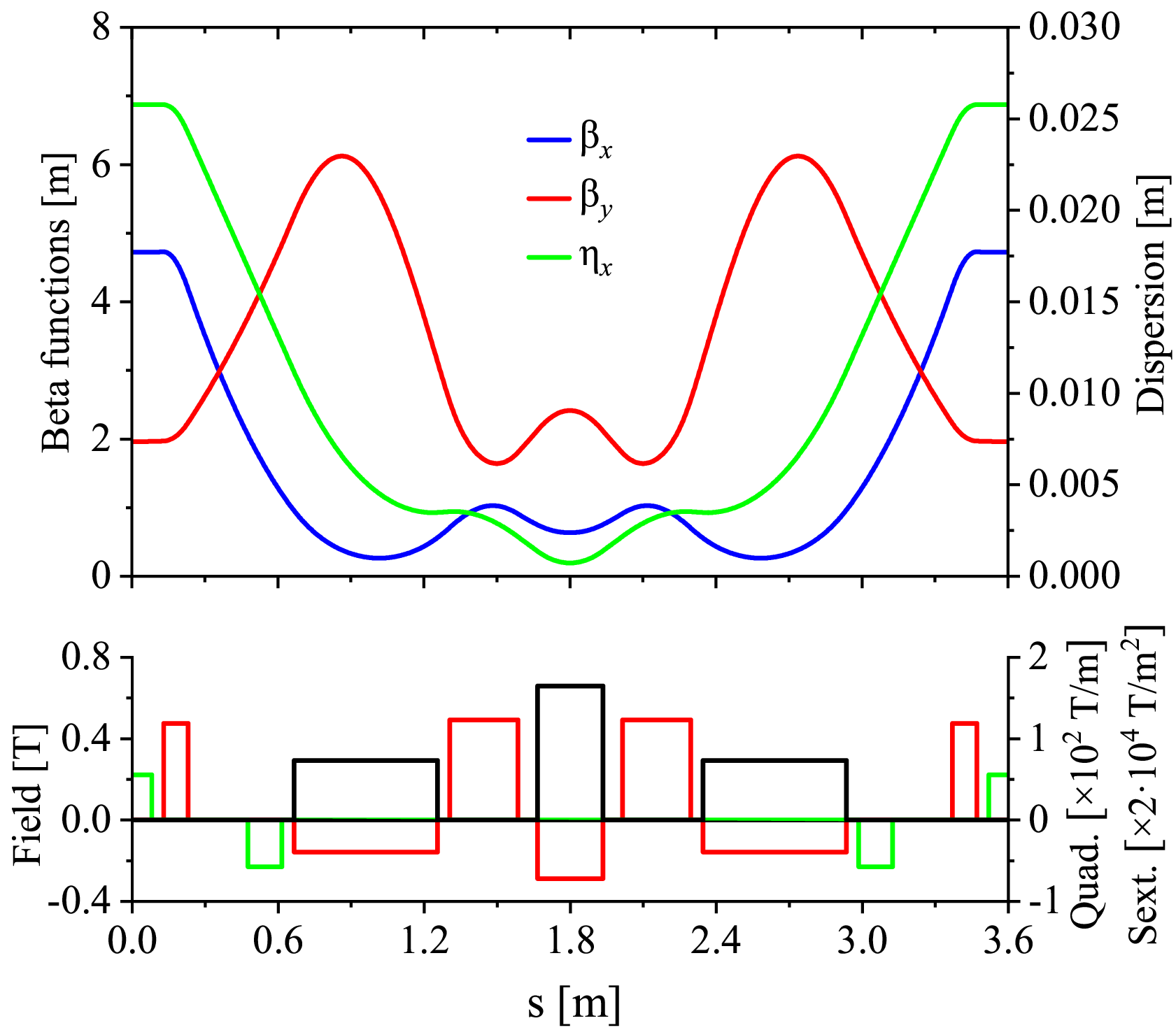}
  \caption{A practical example of a complex unit lattice cell with three main bends. The lower plot shows the distributions of dipole fields (black), quadrupole strengths (red) and sextupole strengths (green) in the cell. Note that the definition of sextupole strengths here is twice that in the OPA code~\cite{streun2017opa}.}
  \label{fig5}
\end{figure}

Now we will present a practical example of a complex unit lattice cell, where the strengths of bends, quadrupoles and sextupoles are all taken into consideration. This complex unit cell has three CBs, and Fig.~\ref{fig5} shows its linear optics, magnet layout and strengths. The central CB with a higher dipole field and the two outer CBs with a lower field form macroscopic-scale longitudinal gradient fields as stated in the previous subsection. The central CB has the lowest dispersion. The bending angle of the complex unit cell is $3^\circ$, and the cell length is 3.6 m. The horizontal and vertical cell tunes are (4/5, 1/5), which will be used for designing a higher-order achromat (HOA) based MBA lattice in the next section. The natural emittance is 14.6 pm$\cdot$rad at 3.0 GeV, and other parameters are listed in Table \ref{table1}.

\begin{table}
\centering
\caption{Parameters of a complex unit cell with three main bends and three identical conventional unit cells.}
\begin{tabular}{lcc}
\hline
 Parameters & Complex cell & 3$\times$Conv. cell  \\\hline
 Energy [GeV] & \multicolumn{2}{c}{3.0} \\
 Length [m] & \multicolumn{2}{c}{3.6} \\
 Bending angle [$^\circ$]& \multicolumn{2}{c}{3.0} \\
 Nat. emittance $\epsilon$ [pm$\cdot$rad] & 14.6 & 15.4 \\
 Betatron tunes (H/V) & (4/5, 1/5) & 3$\times$(4/15, 1/15) \\
 Nat. chromaticities (H/V) & -1.11/-0.90 & -0.72/-0.98 \\
 Mom. comp. factor & 6.1$\times$10$^{-5}$ &  6.2$\times$10$^{-5}$ \\
 Hor. damping partition $J_x$ & 1.84 & 2.09 \\
 $\epsilon \cdot J_x$ [pm$\cdot$rad] & 26.9 & 32.2 \\
 \hline
 \label{table1}
\end{tabular}
\end{table}

\begin{figure} 
  \centering
      \includegraphics[width=0.45\textwidth]{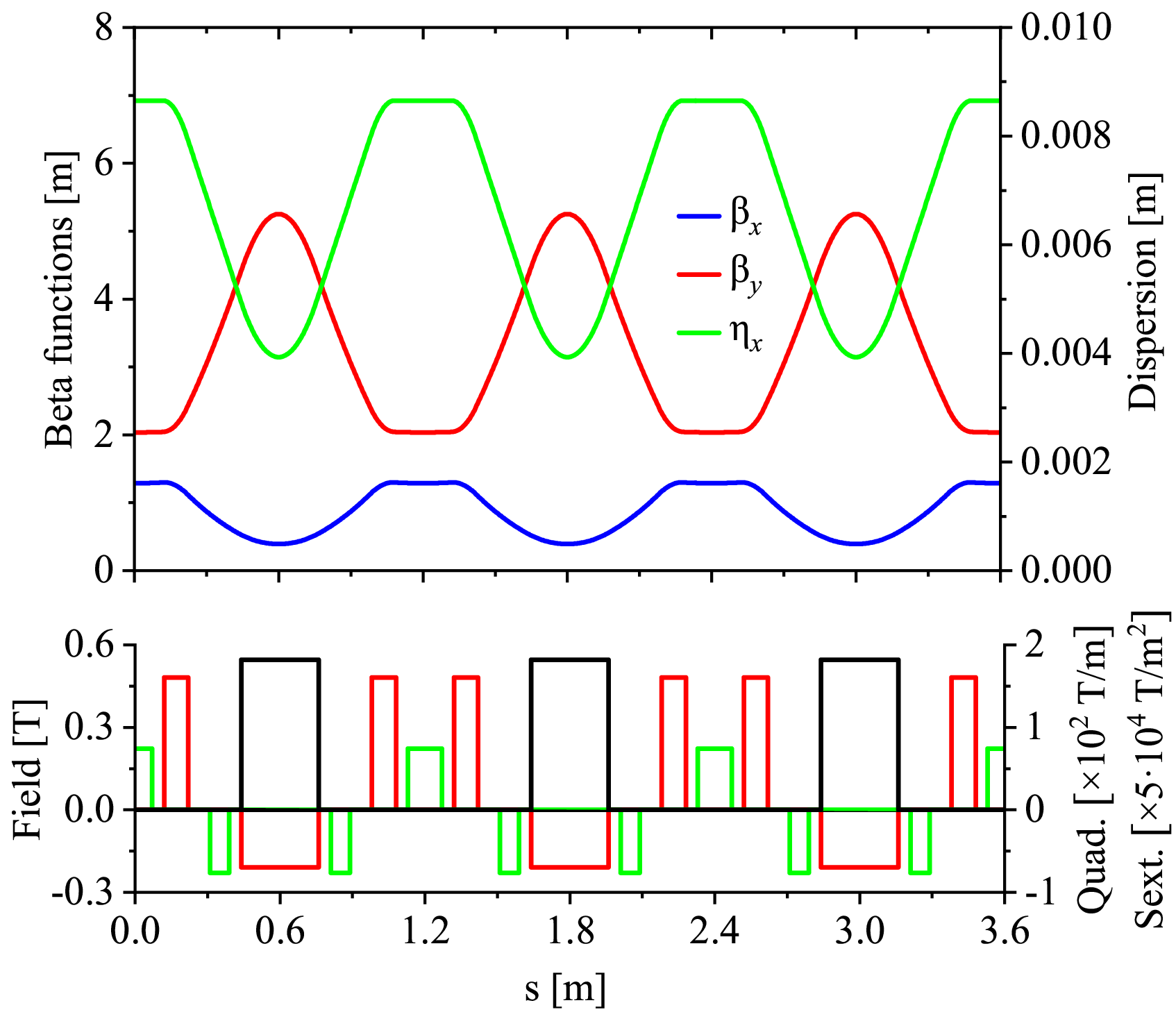}
  \caption{Three identical conventional unit lattice cells. Note that the unit of sextupole strengths here is 2.5 times that in Fig.~\ref{fig5}.}
  \label{fig6}
\end{figure}

For comparison, Fig.~\ref{fig6} shows three identical conventional unit cells, also with CBs as main bends, and the parameters are also presented in Table \ref{table1}. The total angle, length and tunes of these three conventional unit cells are the same as those of the complex unit cell in Fig.~\ref{fig5}. The betatron tunes of a conventional unit cell, (4/15, 1/15), were also used in an HOA based 18BA lattice~\cite{Bengtsson2019Towards}. Compared to the conventional cell, the complex cell has a slightly lower natural emittance, while its normalized emittance, $\epsilon \cdot J_x$, is obviously lower due to smaller $J_x$. So, for the complex cell, there is a greater potential to further reduce the natural emittance by increasing $J_x$ using RBs. In contrast, the $J_x$ of the conventional cell is already quite large, exceeding 2, as shown in Table \ref{table1}.

The maximum dispersion of the complex cell is about 3 times that of the conventional cell, smaller than the factor of 4$\sim$5 in the simplified model. If the three CBs do not have macroscopic-scale longitudinal gradient fields, the maximum dispersion will be reduced. However, the sum of natural chromaticities of the complex cell is larger than that of the three conventional cells, which is not favorable for reducing sextupole strengths. The strengths of magnets are shown in the lower plots of Fig.~\ref{fig5} and Fig.~\ref{fig6}. For the sextupoles, their strengths were calculated with chromaticities corrected to (0, 0). The integrated strengths of sextupoles in the complex cell are only about 1/3$\sim$1/2 of those in the conventional cell. In the complex unit cell, the quadrupoles in the central part have stronger integrated strength, but we can increase their lengths to reduce the strength due to the availability of more space. The strengths of quadrupoles in the complex cell are about 70\% of those in the conventional cell. Considering the strengths of all magnets, the complex cell in Fig.~\ref{fig5} can have a larger vacuum chamber compared to the conventional cell in Fig.~\ref{fig6} due to weaker magnet strengths.

\section{MBA Lattice with Complex Unit Cells}

\begin{figure*}
\includegraphics[width=0.85\linewidth]{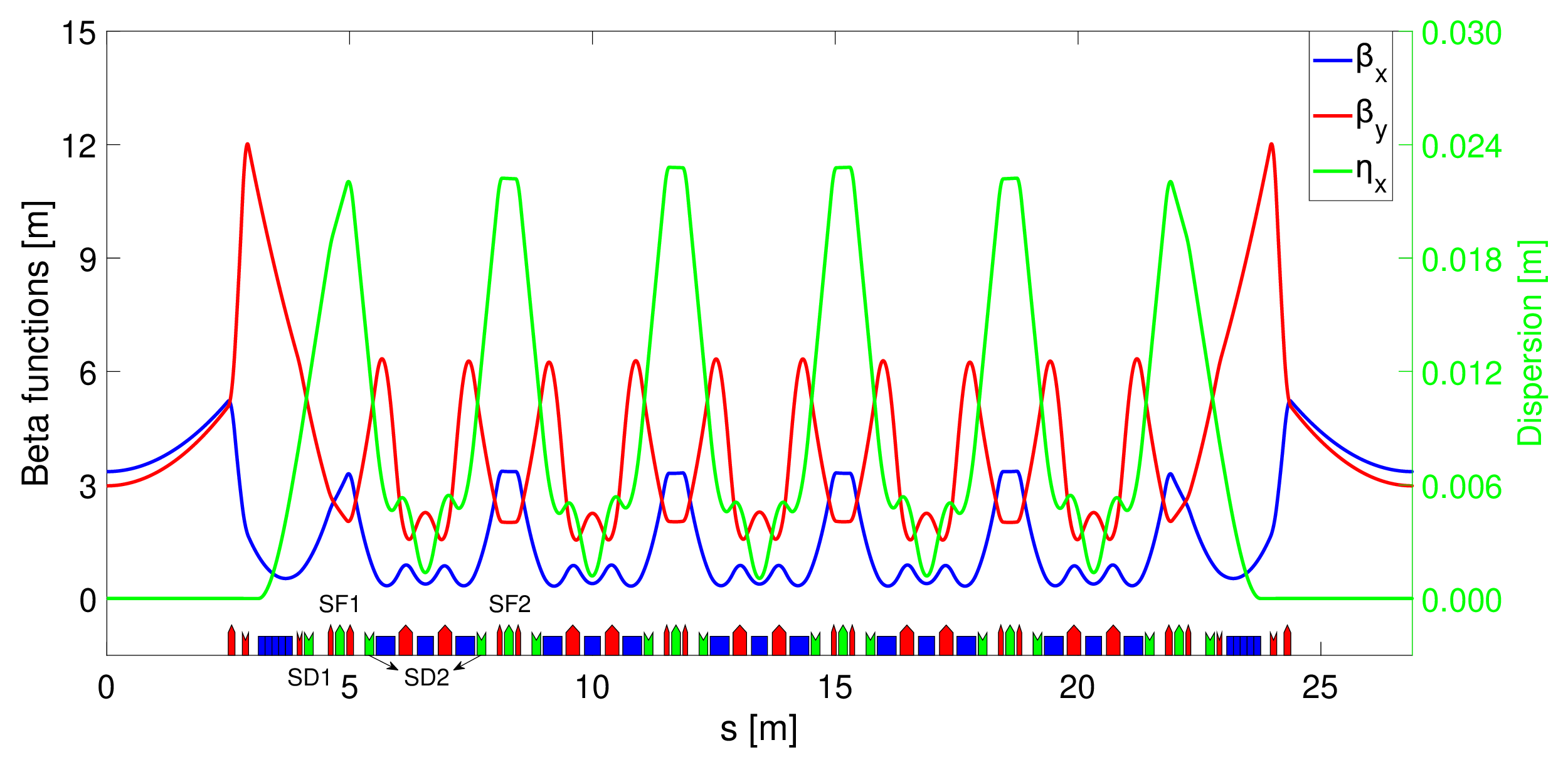}
\caption{\label{fig7} Optical functions and magnet layout of the 17BA lattice with complex unit cells. In the magnet layout, bends are shown in blue, quadrupoles in red and sextupoles in green.}
\end{figure*}

We further designed an MBA lattice with complex unit cells with an emittance goal on the order of 10 pm$\cdot$rad at 3.0 GeV. The complex unit cells use a similar magnet layout as in Fig.~\ref{fig5}. The designed storage ring has a circumference of 537.6 m and consists of 20 identical lattice cells. To mitigate the nonlinear effects, the HOA concept was adopted in the lattice design. The designed lattice is shown in Fig.~\ref{fig7}, which is a 17BA with 5 identical complex unit cells (denoted as 17BA-1), and each unit cell has horizontal and vertical tunes of (4/5, 1/5). The main storage ring parameters are shown in Table \ref{table2}. The natural emittance is 19.3 pm$\cdot$rad.

\begin{figure*}
\includegraphics[width=0.85\linewidth]{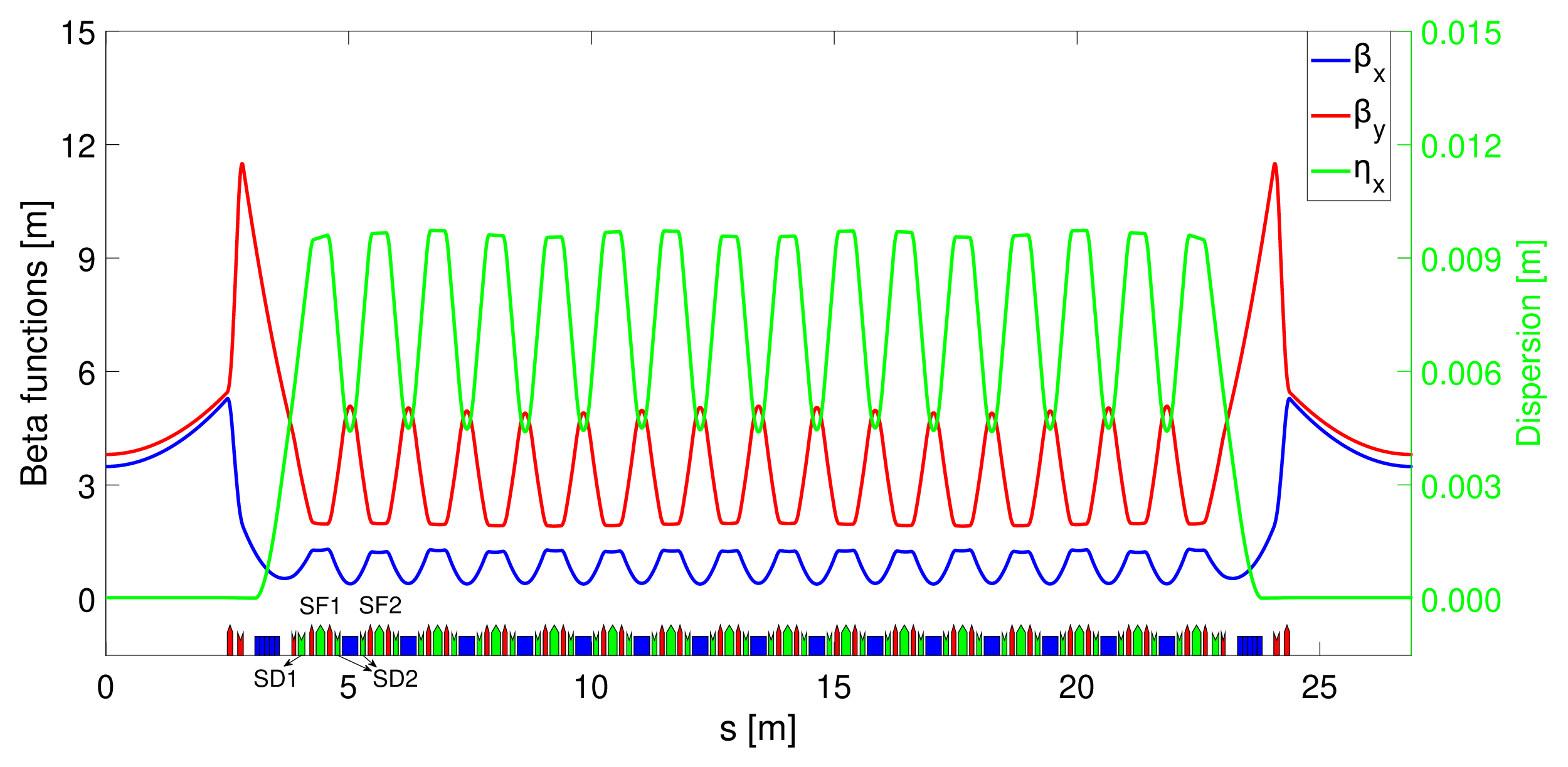}
\caption{\label{fig8} Optical functions and magnet layout of the 17BA lattice with conventional unit cells.}
\end{figure*}

For comparison, a 17BA lattice with conventional unit cells (denoted as 17BA-2) was also designed for the same storage ring, which is shown in Fig.~\ref{fig8}. The unit cells follow a similar magnet layout as in Fig.~\ref{fig6}, and the tunes of each unit cell are (4/15, 1/15) to make an HOA. The main storage ring parameters are also shown in Table \ref{table2}, and the natural emittance is 21.7 pm$\cdot$rad with a larger $J_x$.

\begin{table}
\centering
\caption{Main storage ring parameters of two 17BA lattices with complex unit cells (17BA-1) and conventional unit cells (17BA-2).}
\begin{tabular}{lcc}
\hline
 Parameters   &     17BA-1   &   17BA-2  \\\hline
 Energy [GeV] & \multicolumn{2}{c}{3.0} \\
 Circumference [m] & \multicolumn{2}{c}{537.6} \\
 Number of cells & \multicolumn{2}{c}{20} \\
 Nat. emittance [pm$\cdot$rad] & 19.3 & 21.9 \\
 Betatron tunes (H/V) & 96.14/27.14 & 96.14/27.14 \\
 Nat. chromaticities (H/V) & -110.5/-128.6 & -92.7/-123.0 \\
 Mom. comp. factor & 4.4$\times$10$^{-5}$ &  5.4$\times$10$^{-5}$ \\
 Damping partition \\ numbers (H/V/L) &  1.85/1.0/1.15 & 2.12/1.0/0.88 \\
 \hline
 \label{table2}
\end{tabular}
\end{table}

From Figs.~\ref{fig7} and~\ref{fig8}, we see that the maximum dispersion of 17BA-1 is significantly higher than that of 17BA-2. But the difference in the maximum dispersion between these two 17BA lattices is smaller than that between the two kinds of unit cells in Figs.~\ref{fig5} and~\ref{fig6}. This is because: 1) the matching section of 17BA-1 is longer than that of 17BA-2, and thus the length of the complex unit cell of 17BA-1 is shorter than that of three conventional unit cells of 17BA-2; 2) in addition to reducing the sextupole strengths, we also considered weakening the nonlinear dynamics effects in the lattice design. From Figs.~\ref{fig7} and~\ref{fig8}, we can also see that 17BA-1 has more free space in the arc section, which facilitates the installation of diagnostics and vacuum components.

\begin{figure} 
  \centering
      \includegraphics[width=0.45\textwidth]{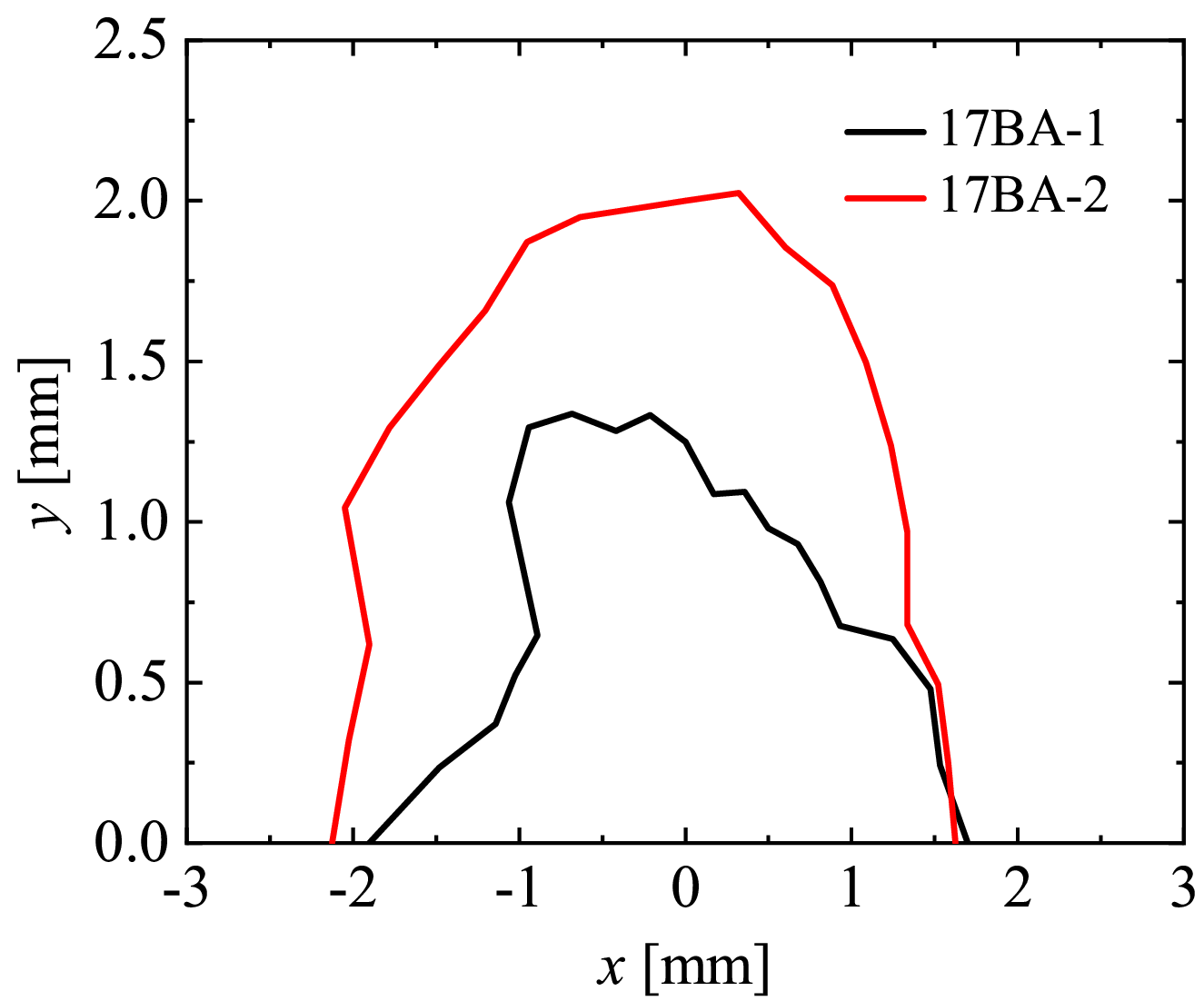}
  \caption{On-momentum DAs of the two 17BA lattices, tracked at the middle of their straight sections.}
  \label{fig9}
\end{figure}

\begin{figure} 
  \centering
      \includegraphics[width=0.45\textwidth]{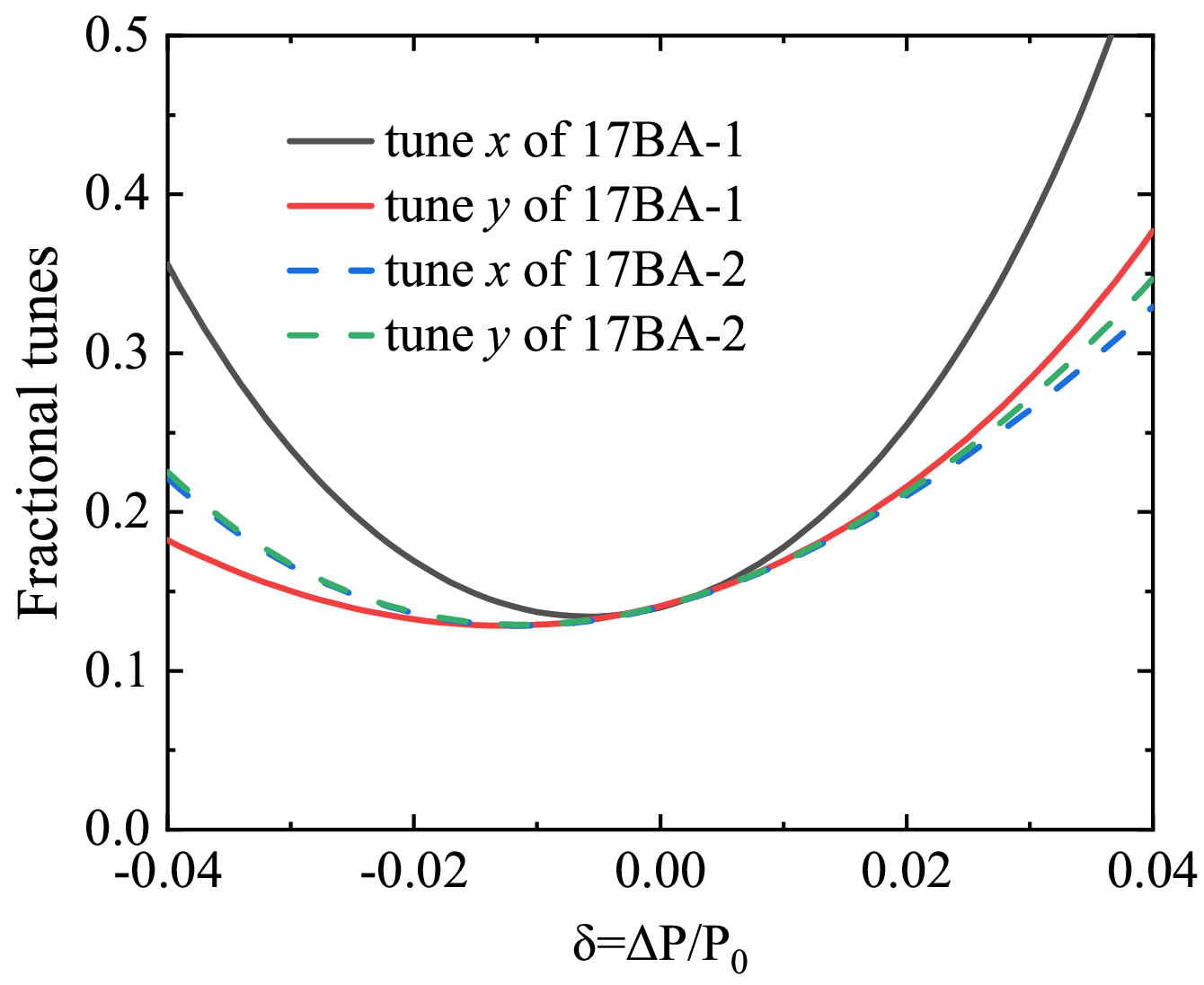}
  \caption{Momentum dependent tune shifts of the two 17BA lattices.}
  \label{fig10}
\end{figure}

In the preliminary nonlinear optimization of the two 17BA lattices, four families of sextupoles were employed with two in the matching sections (SD1 and SF1) and two in the unit cell sections (SD2 and SF2), as shown in the magnet layouts of Figs.~\ref{fig7} and~\ref{fig8}. The OPA code~\cite{streun2017opa} was used to optimize the nonlinear dynamics, and the horizontal and vertical chromaticities were corrected to (2, 2). The strengths of sextupoles of the two lattices are listed in Table \ref{table3}, and we see that the integrated strengths of sextupoles of 17BA-1 are about 2$\sim$3 times lower than those of 17BA-2. The optimized dynamic apertures (DAs) of the two lattices are shown in Fig.~\ref{fig9}. The horizontal DAs are about 1.5 or 2 mm. The vertical DA of 17BA-1 is smaller, but its vertical beta function at the middle of straight section is also somewhat smaller. For the 19BA lattice in Ref.~\cite{Tavares2018future} and the 18BA lattice in Ref.~\cite{Bengtsson2019Towards}, their horizontal DAs are also about 1.5 mm. Figure~\ref{fig10} shows the momentum dependent tune shifts. Although 17BA-1 has larger tune shifts than 17BA-2, its tunes do not cross half-integer resonance lines within 3.5$\%$ momentum deviation. Therefore, although the strengths of sextupoles of 17BA-1 are much weaker compared to 17BA-2, its nonlinear dynamics performance is not better than that of 17BA-2. To further improve the nonlinear dynamics, we can group the sextupoles into more families and consider using octupoles, but here we will not study it.

\begin{table}
\centering
\caption{Parameters of sextupoles of the two 17BA lattices. The definition of sextupole strengths here is twice that in the OPA code.}
\begin{tabular}{lccc}
\hline
Sextupole families  &  Lengths  &  Strengths  &  Integrated strengths  \\ 
            & [m]      &  [T/m$^2$] &  [T/m]  \\ \hline
17BA-1: SD1  & 0.16     & -10312    &  -1650 \\
17BA-1: SF1  & 0.18     & 11618      &  2091  \\
17BA-1: SD2  & 0.16     & -14802     &  -2368 \\
17BA-1: SF2  & 0.18     & 15037      &  2707  \\

17BA-2: SD1  &  0.14    &  -23552    &  -3297  \\
17BA-2: SF1  &  0.18    &  37889     &  6820 \\ 
17BA-2: SD2  &  0.10    &  -38305    &  -3831  \\
17BA-2: SF2  &  0.18    &  36812     &  6626 \\ \hline
 \label{table3}
\end{tabular}
\end{table}

Revisiting Fig.~\ref{fig7}, if the three bends of a complex unit cell are seen as three pieces of a single bend, then 17BA-1 can be seen as a 7BA lattice. MBA lattices can be classified into two types: one is the MBA lattice with distributed chromatic correction, such as 17BA-2, and the other is the hybrid MBA lattice. For 17BA-1, it can be seen as an MBA lattice between the two types, and we can call this new type the MBA lattice with semi-distributed chromatic correction.

\section{Further Study}

\begin{figure} 
  \centering
      \includegraphics[width=0.45\textwidth]{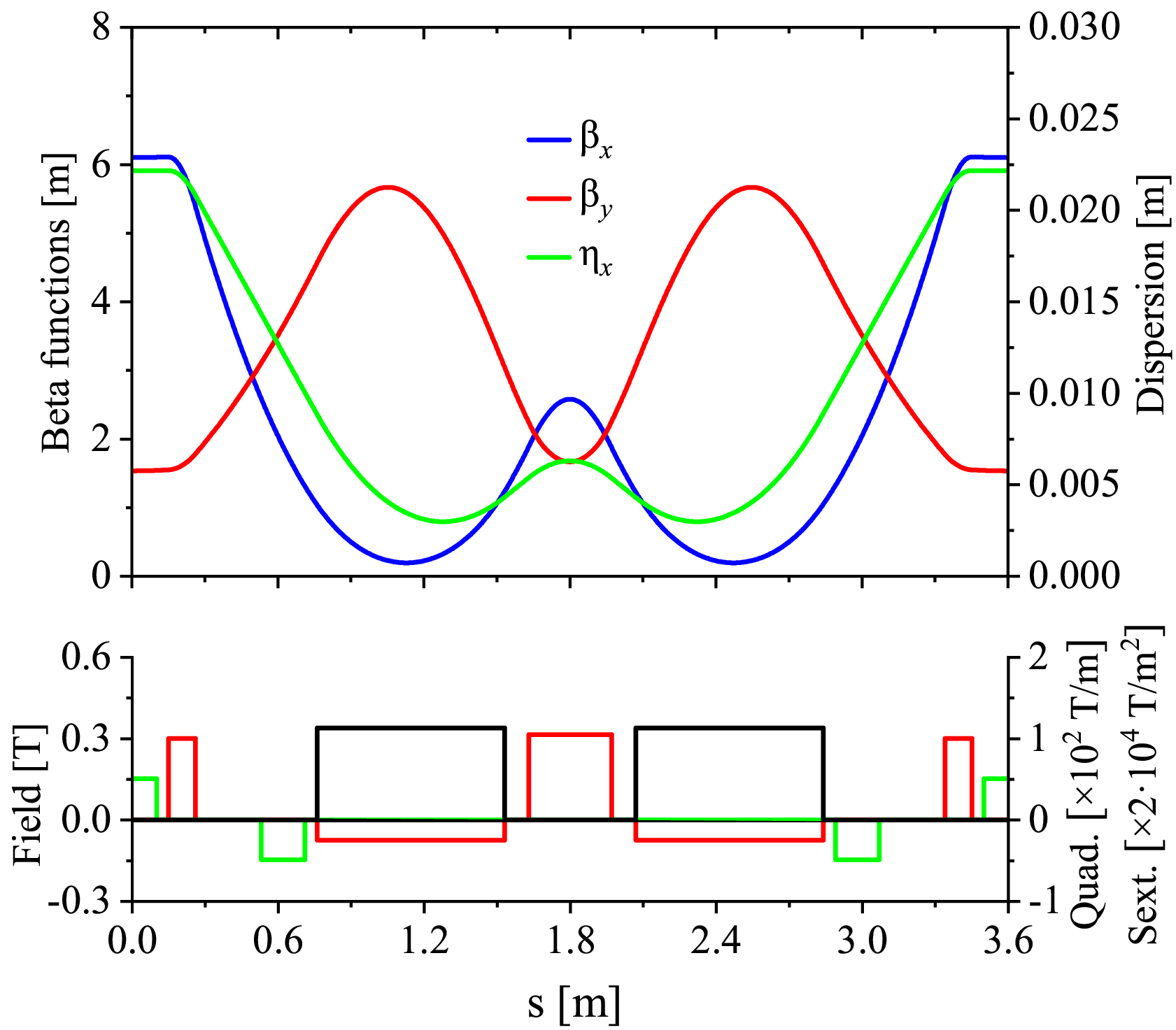}
  \caption{A complex unit lattice cell with two main bends, which has the same energy, length, bending angle and tunes as those of the complex unit cell with three main bends shown in Fig.~\ref{fig5}.}
  \label{fig11}
\end{figure}

As in Section II C, we designed a similar complex unit lattice cell with two main bends, but with the same tunes (4/5, 1/5) and the same length 3.6 m as the previous complex unit cell in Fig.~\ref{fig5}, which is shown in Fig.~\ref{fig11}. For this complex unit cell, the horizontal focusing of bend cell, which can be characterized by the horizontal tune divided by the number of bends, is stronger than that of the previous complex unit cell, and there are fewer magnets. With the same energy and cell bending angle, the natural emittance is 15.3 pm$\cdot$rad with $J_x$=1.64. The horizontal and vertical natural chromaticities are (-1.54, -0.69), the sum of which is larger than that of the previous complex unit cell. With chromaticities corrected to (0, 0), the integrated strengths of sextupoles are about 10\% higher than those of the previous complex unit cell. However, due to fewer magnets used and thus more available space, we can increase the lengths of magnets to reduce their strengths. As shown in Fig.~\ref{fig11}, the strengths of magnets of this complex unit cell are lower than those of the previous complex unit cell. And the longest drift in this complex unit cell is also longer.

Therefore, for two complex unit cells with the same tunes and length, one with fewer bends but stronger focusing and the other with more bends but weaker focusing, it is preferable to choose the former, in order to reduce magnet strengths and save space. Besides, the nonlinear dynamics performance also needs to be considered in the lattice design. The comparison of nonlinear dynamics between these two kinds of complex unit cells will not be studied here.

\section{Conclusion}
The unit lattice cell, the basic component of a storage ring lattice, has been intensively studied over the past decades, from the TME unit cell~\cite{teng1984minimizing} to the LGB/RB unit cell~\cite{Riemann2019low}. In past studies, the main focus has been on reducing the emittance of unit cell. Storage ring light sources are being developed toward the true diffraction-limited emittance. The most ambitious lattice design would be the 19BA lattice for the future upgrade plan of MAX IV. In such a lattice, the strengths of magnets are extremely strong and the space between magnets is also very limited, which can pose significant challenges in storage ring physics and technology. In order to address such an issue, in this paper we proposed the concept of complex unit cell. Different from the conventional unit cell with only one main bend, the complex unit cell has more than one main bend in the central part, and the central part shares the sextupoles located at both sides for chromaticity correction. And different from past studies of unit cell, in our study the main focus is on reducing magnet strengths and saving space. 

The complex unit cell was first studied with a simplified model and then practically designed and compared with the conventional unit cell. As the number of main bends in the central part increases, the integrated strengths of sextupoles can be reduced; and when the number of main bends is larger than two, the dipole fields of main bends tend to form macroscopic-scale longitudinal gradient field in order to reduce sextupole strengths. Compared to the conventional unit cell, the magnet strengths of the complex unit cell can be lower and also more space can be left at both sides of the cell. The difference in emittance between the complex and conventional unit cells is very small, but the former has smaller $J_x$, indicating that its emittance can be further reduced more effectively by increasing $J_x$ using RBs. In the further study, it was also found that it is preferable to use the complex unit cell with stronger focusing for better reducing magnet strengths and saving space. For the nonlinear dynamics, it is not the focus of the study in this paper. Besides, the designed MBA lattice with complex unit cells can be seen as a new type of MBA lattice, which we call the MBA lattice with semi-distributed chromatic correction.

The complex unit cell explores a new pathway toward reaching the true diffraction-limited emittance but without using so extremely strong magnets and compact magnet layout as in the 19BA lattice for the MAX IV upgrade plan, which could pave the way for the further development of diffraction-limited storage rings with limited circumferences. As in the complex bend where there is no beam position monitor (BPM) and corrector between its small magnets, the central part of the complex unit cell can also be seen as a single magnet and there can be no BPM and corrector with the development of magnet technology.

\begin{acknowledgments}
This work was supported by the National Natural Science Foundation of China under Grant Nos. 11875259 and 12075239.
\end{acknowledgments}

\nocite{*}

\bibliography{apssamp.bib}
\end{document}